# ACS APPLIED NANO MATERIALS

www.acsanm.org

Article

# FeCo Nanowire−Strontium Ferrite Powder Composites for Permanent Magnets with High-Energy Products

J. C. Guzmán-Mínguez, S. Ruiz-Gómez, L. M. Vicente-Arche, C. Granados-Miralles, C. Fernández-González, F. Mompeán, M. García-Hernández, S. Erohkin, D. Berkov, D. Mishra, C. de Julián Fernández, J. F. Fernández, L. Pérez, and A. Quesada*

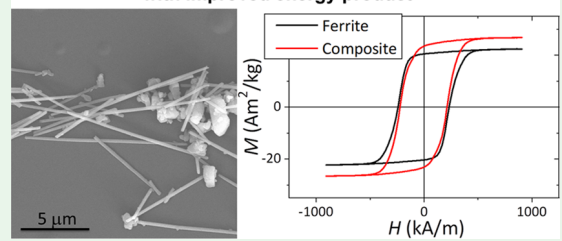

**ABSTRACT:** Due to the issues associated with rare-earth elements, there arises a strong need for magnets with properties between those of ferrites and rare-earth magnets that could substitute the latter in selected applications. Here, we produce a high remanent magnetization composite bonded magnet by mixing FeCo nanowire powders with hexaferrite particles. In the first step, metallic nanowires with diameters between 30 and 100 nm and length of at least 2 μm are fabricated by electrodeposition. The oriented as-synthesized nanowires show remanence ratios above 0.76 and coercivities above 199 kA/m and resist core oxidation up to 300 °C due to the existence of a >8 nm thin oxide passivating shell. In the second step, a composite powder is fabricated by mixing the nanowires with hexaferrite particles. After the optimal nanowire diameter and composite composition are selected, a bonded magnet is produced. The resulting magnet presents a 20% increase in remanence and an enhancement of the energy product of 48% with respect to a pure hexaferrite (strontium ferrite) magnet. These results put nanowire−ferrite composites at the forefront as candidate materials for alternative magnets for substitution of rare earths in applications that operate with moderate magnet performance.

## ■ INTRODUCTION

The best magnets in the world, amounting to 80% of the worldwide sales of the $20 billion permanent magnet market, contain rare-earth elements (REEs).[1] Their properties are far superior, which creates a large gap between the energy product—the figure of merit of a magnet—of REE magnets and of the other families of magnets, such as hard ferrites and alnicos.[2] Unfortunately, as critical raw materials, REEs are associated with supply risk, price volatility, and environmentally harmful extraction and separation.[3] A large number of applications, covering an important slice of the market, only require moderate magnet performance to operate; nevertheless, in the absence of a gap magnet, these technologies have no alternative but to rely on critical REE magnets. In this framework, there is a strong demand for the development of alternative REE-free or REE-lean gap magnets that could effectively substitute REEs in these applications.[4,5]

A considerable amount of projects and initiatives have focused their efforts on this issue in the past few years[2,6] by exploring a relatively wide variety of solutions that include developing completely new magnetic materials,[5] improving the properties of existing magnetic phases,[7−9] reducing the content of heavy REEs—which are the real critical issues—in existing REE magnets,[10] and fabricating composite materials by combining phases with different magnetic properties.[11−13]

One of the main challenges associated with the development of permanent magnets is that it is hard to come across a material in nature that simultaneously hosts a large coercivity and high magnetization.[14] For this reason, shape anisotropy has been often employed as a tool to sustain a certain resistance to demagnetization in high-magnetization systems, for instance, by fabricating high-aspect-ratio transition-metal structures.[15,16] A great deal of work has been focused on Co nanowires (NWs) owing to the relatively large magnetocrystalline anisotropy of Co compared to other magnetic transition metals such as Fe and Ni.[14−18] Promising results[19,20,22] have been obtained in this system, including the fabrication of NW-only dense pellets with improved magnetic properties.[21,23] Fe, FeNi, and FeCo nanowires have been investigated as well.[24−29] They present

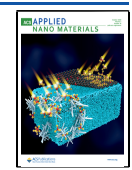

9842

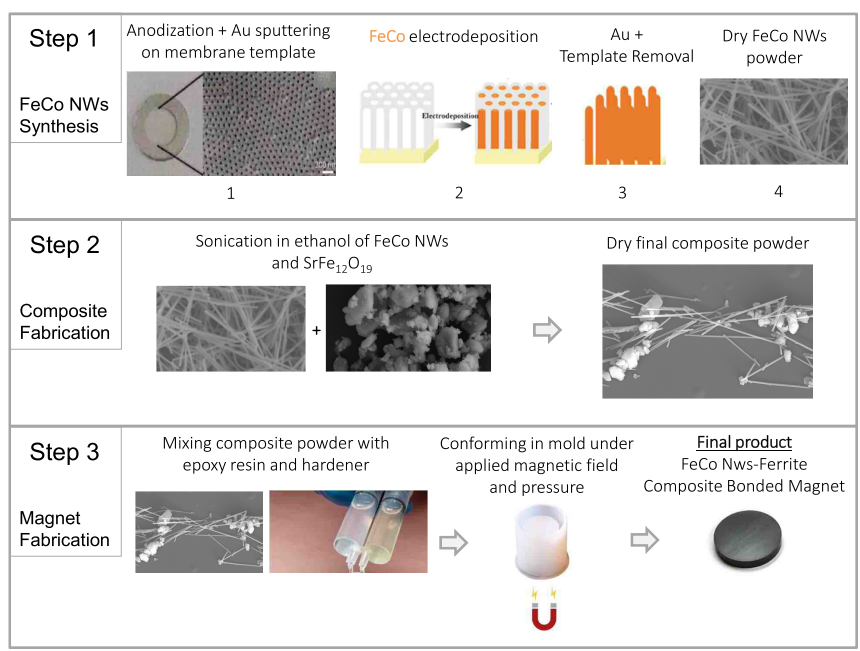

**Figure 1.** Schematic of the fabrication process of the FeCo NWs, the ferrite−FeCo NW composites, and the composite bonded magnet.

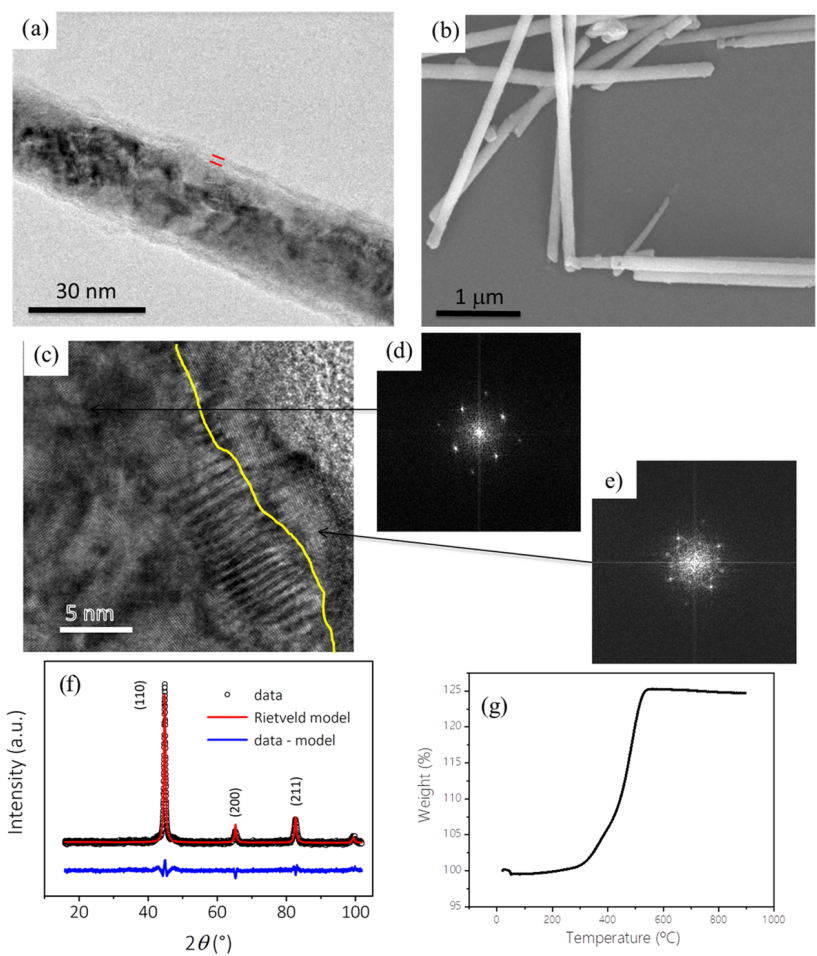

**Figure 2.** (a) TEM image of a 50 nm diameter isolated NW. (b) SEM image of the 50 nm diameter NW powder. (c) HRTEM image of the side of the NW from (a). The yellow line delimits the core−shell frontier. Fast Fourier transform (FFT) patterns of (d) core and (e) shell. (f) XRD pattern and corresponding Rietveld model of the 50 nm NWs and (g) TGA curve from room temperature (RT) to 900 °C measured in air.

larger magnetization and lower coercivity than Co, which makes them suitable for a wide array of application sectors that include biomedical,[30−33] environmental (in this case, ferrite NWs),[34] and sensors,[35] among others. In addition, arrays of low-

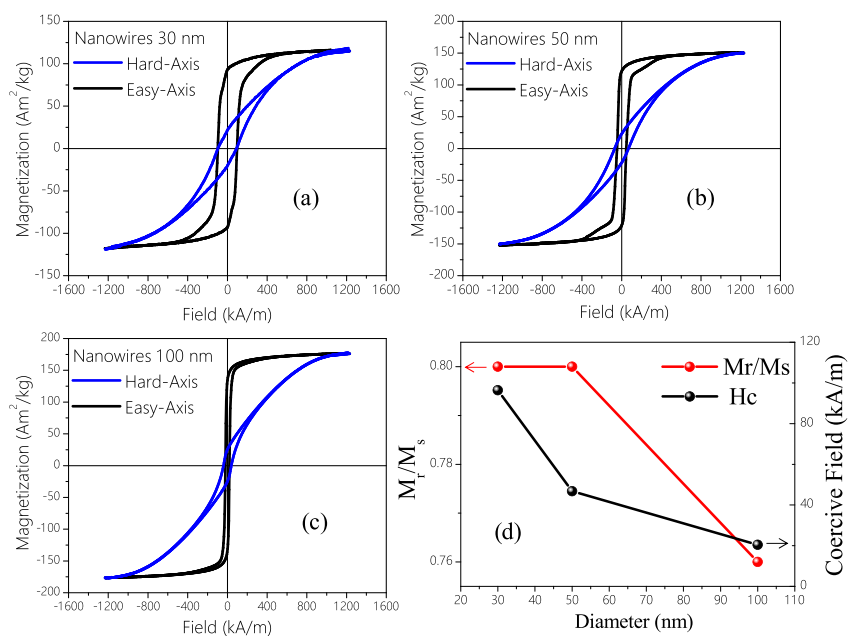

**Figure 3.** Magnetization curves measured at RT of (a) 30 nm, (b) 50 nm, and (c) 100 nm diameter NW powders oriented in an external magnetic field inside a bonding glass matrix. Black and blue curves correspond to the applied field being parallel (black) and perpendicular (blue) to the nanowire long axis. (d) Evolution of the remanence-to-saturation ratio ($M_r/M_s$) and coercivity ($H_c$) as a function of the NW diameter.

coercivity NWs are excellent systems to study and analyze the complex hysteretic and reversal mechanisms of ferromagnets.[33,36−39]

In view of their potential application as permanent magnets and to minimize the low coercivity problem, they have been used as a secondary phase in composite systems together with harder magnetic materials, with reports mainly focused on MnBi as the host phase.[40−42] Again, promising results have been achieved in these systems, showing the potential of magnetic nanowires as building blocks for permanent magnets. However, the production yield of these nanostructures is usually low and the associated fabrication costs are high, which explains their absence in commercial products nowadays.

In an effort to circumvent these issues, in this work, we focus on the development of composite magnets based on FeCo NWs and strontium ferrite ($SrFe_{12}O_{19}$), as ferrites are known for their availability and low price ($1.5/kg). In the first part of the work, the experimental and theoretical study of the structures and properties of electrodeposited NWs as a function of their diameter is presented by means of electron microscopy, X-ray diffraction (XRD), thermogravimetric analysis (TGA), and magnetometry. Micromagnetic simulations were performed to understand the dimension dependence of the magnetic properties of the FeCo NWs and to determine the magnetic structure. High-magnetization FeCo NWs of different diameters were mixed and oriented with different concentrations of hard hexaferrite micropowders. After selecting an optimal diameter, the most adequate mixture composition is decided upon by characterizing powder composite samples of a few milligrams. A bulk bonded composite magnet (0.3 g) is fabricated out of the optimal composition and compared to a pure ferrite reference magnet fabricated and characterized under identical conditions. Figure 1 presents a schematic of the experimental fabrication process of the FeCo NWs, the composites, and the bonded magnet.

## ■ RESULTS

The morphological and structural properties of the FeCo NW dry powder were characterized by transmission and scanning electron microscopy (TEM and SEM) and X-ray diffraction (XRD). Figure 2a shows the TEM images of a FeCo nanowire grown in the 50 nm pore template. First, the NW diameter is measured to be $D_{50nm}$ = 49 nm. The darker core of the nanowire is observed to be surrounded by a thin layer that appears slightly brighter in the image. This is consistent with the formation of a passivating oxide layer on the surface of the FeCo NWs, resulting in a core−shell type of structure that protects the metallic core from further oxidation.[43] The diameter of the metallic core is $D_{core}$ = 35 nm. This value is in agreement with previous reports,[43,44] as a shell with a thickness of 7 nm is inferred by subtracting outer and inner diameters. The SEM image in Figure 2b shows that all FeCo NWs obtained in the batch present a diameter close to 50 nm, confirming a very narrow dispersion of values. In addition, an average length of 6 μm is observed. The 30 and 100 nm diameter powders were analyzed by SEM as well, as shown in Figure S1 of the Supporting Information (SI), confirming a narrow dispersion of diameters. For the 100 nm FeCo NWs, an average length of 2 μm is observed.

Figure 2c shows a high-resolution TEM (HRTEM) image of the 50 nm FeCo NW. Lattice fringes can be observed, although crystallite sizes and lattice constants are not easily extractable. Two fast Fourier transform (FFT) patterns are shown in Figure 2d,e. A body-centered cubic (bcc) structure is revealed at the core, as expected for metallic FeCo alloys. The pattern of the shell is significantly noisier and no structure is easily identifiable, although it is clearly different from the core FFT pattern.

The XRD pattern of the powder in Figure 2d only shows diffraction maxima at the (110), (200), and (211) Bragg positions corresponding to a bcc lattice, with a lattice parameter of 2.856 Å. Using the equation in ref 45, from the lattice parameter, we infer a composition $Fe_{49}Co_{51}$, with no detection of other phases up to the resolution limit of the instrument (approximately 5 wt %) and that the Fe and Co atoms are evenly distributed among the sites of the bcc cell. The average crystallite diameter obtained from the Rietveld model is approx. 12 nm. This implies that the depth of a NW comprises several crystallites that may explain the difficulties in observing clear features in the HRTEM image. The lack of other diffraction maxima, in particular, related to FeCo oxides, suggests a poor crystallinity/amorphous structure of the oxide shell layer, as already indicated by the diffused FFT spots in Figure 2e.

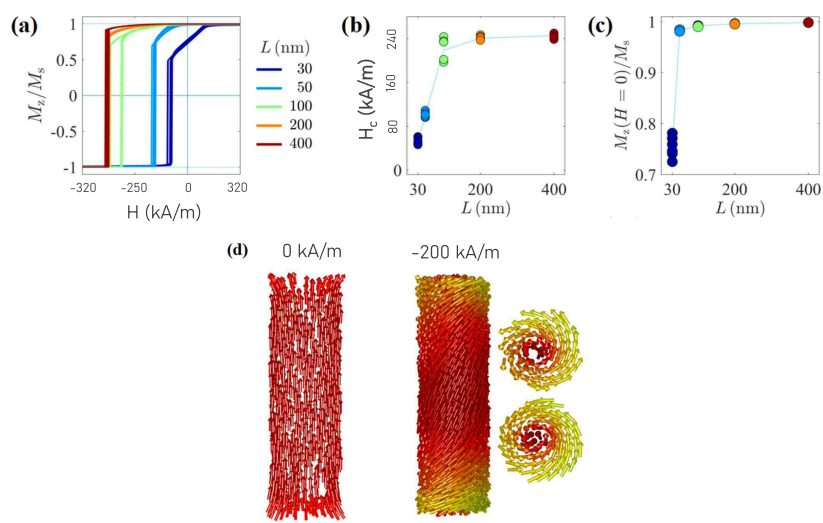

**Figure 4.** Results of micromagnetic simulations: (a) hysteresis curves, (b) coercivity $H_c$, and (c) remanence ratio along the long axis ($z$ direction) $M_z/M_s$ as a function of the length of a NW with $D$ = 30 nm. Curves and dots of the same color present different geometrical realizations of the finite mesh element structure. (d) Spin structure of the NW with $D$ = 30 nm and $L$ = 100 nm at remanence (0.0 T) and −200 kA/m negative field. For a better visibility, only magnetic moments for the mesh elements on the surface are shown.

The TGA curve in Figure 2e shows a very mild weight gain from RT to 300 °C, a temperature at which a more pronounced weight gain commences, reaching a maximum value of 25.3% at 520 °C and subsequently staying relatively constant up to 900 °C. Previous studies allow us to correlate this TGA curve with the oxidation of the FeCo bcc alloy nanoparticles to $CoFe_2O_4$ (Co ferrite) under high-temperature annealing in air.[43,46] Assuming that is the case here if the FeCo NWs were 100% metallic, a 37% weight gain should be observed during the oxidation to Co ferrite. From the values of the length and outer/core diameter of the 50 nm NW, the volume of the metallic core in an average FeCo NW is calculated to be $V_{core}$ = 24.4 × 10$^6$ nm$^3$, while that of the oxide shell is $V_{shell}$ = 20.5 × 10$^6$ nm$^3$. Using the density of FeCo of 8.1 g/cm$^3$ and that of Co ferrite of 5.23 g/cm$^3$,[47] we calculate the mass of the oxide shell to amount to 28.6% of the total mass of the FeCo NW powder. Thus, since only 71.4% of the sample mass is metallic, the expected weight gain during oxidation of the 50 nm MWs is 26.4% (this number is calculated as the 37% of the metallic part of the sample, i.e., 71.4%, with 37% being the gain weight associated to the oxidation of FeCo alloy to $CoFe_2O_4$). The measured 25.3% weight gain is in good agreement with this value, experimentally supporting that the FeCo NW powders approximately present a 71 wt % FeCo–29 wt % Co ferrite. Moreover, we conclude from the TGA results that the passivating layer protects the metallic core from further oxidation at temperatures up to 300 °C.

The magnetization curves of the three samples, measured in parallel (black) and perpendicular (blue) to the alignment direction, are presented in Figure 3.

The saturation magnetization ($M_s$) values obtained for each diameter are 115, 150, and 176 Am$^2$/kg for 30, 50, and 100 nm, respectively. As a comparison, recent reports found $M_s$ = 240–248 Am$^2$/kg for $Fe_{65}Co_{35}$ NWs,[32,48] although it is important to point out that NWs were fully metallic in those cases wherein the template was not removed and the NWs were not exposed to air. The $M_s$ value for 50 nm FeCo NWs is in agreement with the fact that the sample is composed of 71 wt % FeCo, which is known to have $M_s$ ≈ 210–220 Am$^2$/kg.[44,49] The Co–Fe oxide shell thus has a small contribution to the overall $M_s$. This can be explained by the fact that, on the one hand, Co ferrite with a 1:1 Fe/Co ratio has a significantly smaller $M_s$, around 40–50 Am$^2$/kg, than stoichiometric $CoFe_2O_4$, around 80 Am$^2$/kg.[50] On the other hand, the spin disorder due to low crystallinity, inferred from the TEM and XRD data, further reduces that value. To study the long-term stability of the 50 nm NWs, we remeasured the magnetization curves 1 year after drying the powder and first exposure to air. No significant changes were observed (see Figure S3 of the SI), from which it can be concluded that the oxide shell prevents long-term oxidation (up to 1 year at least) as well. For the 30/100 nm FeCo NWs, the $M_s$ values imply a 52/78 wt % of metallic phase in the nanowires, which is in agreement, following the same calculation executed above, with a passivating shell of thickness $t$ ≈ 6–8 nm, the same as for the 50 nm case.

Notable differences can be observed between the parallel and perpendicular magnetization curves, particularly in the remanence ($M_r$) values, which demonstrate that the samples are significantly anisotropic—remanence-to-saturation ratios ($M_r/M_s$) between 0.8 and 0.76 are observed as shown in Figure 3d—and have indeed been aligned inside the bonding glass. For instance, $M_r$ = 122 Am$^2$/kg for the 50 nm FeCo NWs. It is important to point that, since we are measuring ensembles of NWs and demagnetizing fields exist in the samples, the $M_r$ values measured constitute a lower limit for the real remanence of a single FeCo NW of a given diameter.

Given that FeCo is a material with relatively low magnetocrystalline anisotropy,[46] it is reasonable to assume that the shape anisotropy of the wires is responsible for this behavior, supporting a large $M_r$ along the NW long axis. This picture suggests that the FeCo NWs are mainly in a single-domain

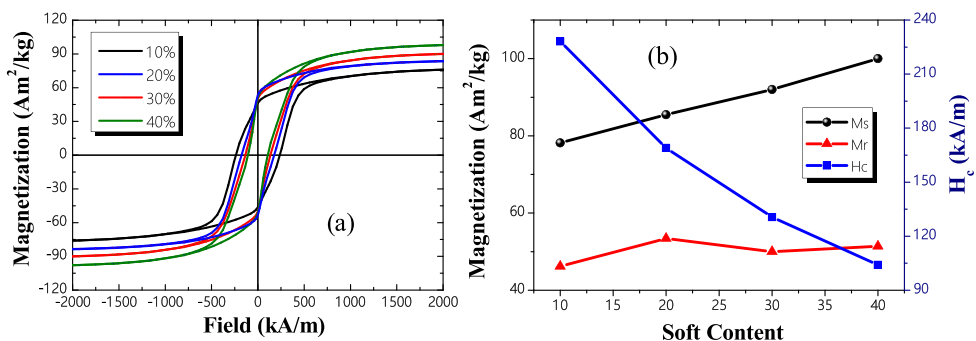

**Figure 5.** (a) Magnetization curve of nonoriented NW (50 nm)−ferrite composite powders with different NW wt %. (b) Coercivity ($H_c$), saturation magnetization ($M_s$), and remanence ($M_r$) as a function of the soft content in the ferrite−NW (50 nm) composites.

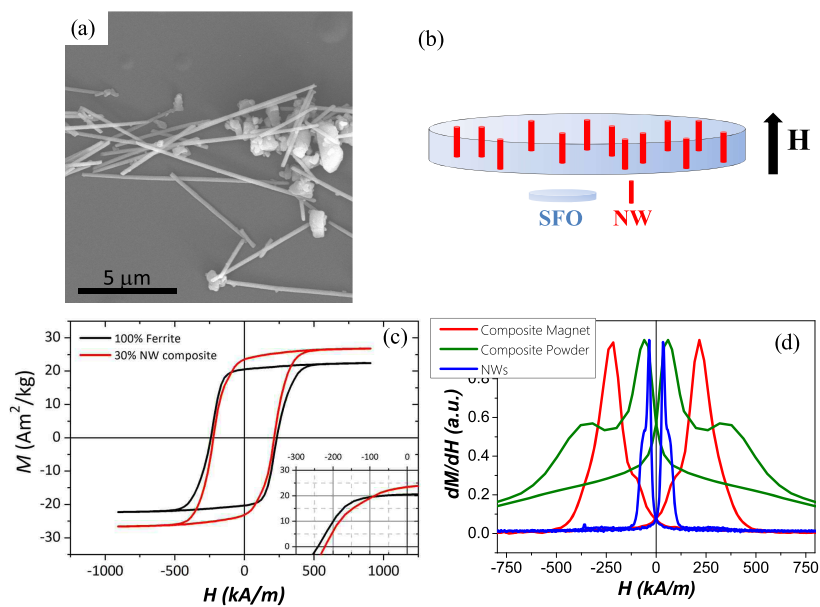

**Figure 6.** (a) SEM image of the 70 wt % strontium ferrite−30 wt % FeCo NW powder. (b) Schematic of the bonded magnet fabricated out of the composite powder, showing NWs (in red) dispersed inside the ferrite magnet (in blue). (c) Magnetization curves at RT of the composite magnet (red) and the reference pure SrM magnet (black). The inset shows the second quadrant. (d) Derivative of the magnetization vs applied field for the isolated 50 nm NWs (blue), the 30−70 wt % composite powder (green), and the composite bonded magnet (red); the latter two are composed of SFO and 50 nm NWs.

magnetic state at remanence, with most of the spins pointing along the long axis.

Moreover, a decrease of coercivity ($H_c$) with FeCo NW diameter is observed, from 96 kA/m for 30 nm NW to 20 kA/m for 100 nm. Recently, coercivities between 55 and 60 kA/m have been reported for 30 nm FeCo NWs (with no oxide shell).[38,48] For 50 nm diameter, a kink in the easy-axis curve is observed in the first and third quadrants. This is likely due to the presence of a secondary phase with higher $H_c$. A small amount of Co ferrite at the shell and/or metallic Co at the core could explain the observation. Coercivity is in any case considerably higher than the normal values for FeCo isotropic particles—$H_c \approx 8$ kA/m.[44] It is true that, assuming a diameter-independent passivating layer thickness, as will be evidenced later in the manuscript, the oxide concentration increases as the diameter decreases, and Co ferrite is known to present a large $H_c$. However, for the 1:1 Fe/Co ratio of our samples, the coercivity of Co ferrite is below 10 kA/m.[50] Thus, it seems more likely that this behavior is explained by an increase in shape anisotropy as the aspect ratio increases and diameter decreases.[15]

To understand the observed high remanence and coercivity values, micromagnetic simulations were performed. Figure 4a−c presents the hysteresis loops, $H_c$, and remanence ratio obtained from the micromagnetic study on FeCo NWs with diameter $D$ = 30 nm and various lengths $L$. In terms of magnetic performance, hysteresis curves (Figure 4a) and corresponding coercivities (Figure 4b) suggest that for $L$ = 100 nm, $H_c$ achieves 83% of its maximum value, while the remanence (Figure 4c) is within a few percentage points of the saturation magnetization starting for $L \geq 50$ nm. The considerably higher remanence ratios obtained in the simulations, compared to those experimentally measured in Figure 3, are due to the fact that the simulations are performed for single NWs and thus in the absence of interaction/demagnetizing fields. For $L$ = 200 nm, i.e., an aspect ratio of 7, the maximum $H_c$ and $M_r$ are already attained.

The details of the spin structure within the wire at remanence and −200 kA/m negative field are shown in Figure 4d for a FeCo NW with $D$ = 30 nm and $L$ = 100 nm. The typical "flower" magnetization distribution at remanence demonstrates an almost single-domain state along the wire axis, with magnetic moments slightly deviating from this axis on top and bottom edges. The spin structure at −200 kA/m is presented by the lateral and top views of both NW edges. The spins on the wire surface deviate from the long axis, forming vortices at the top and bottom edges, as observed in the top view. The total magnetization at these fields suggests that spins of the central core part of the wire do not deviate from the long axis.

Table 1. Magnetic Properties of the Composite Bonded Magnet and a Pure Ferrite Reference Magnet

| oriented samples | $H_c$ (kA/m) | $M_r$ (Am$^2$/kg) | $M_s$ (Am$^2$/kg) | $BH_{max}$ (kJ/m$^3$) |
|---|---|---|---|---|
| composite magnet (30 wt % NW−70 wt % SrM) | 227 | 24 | 28 | 0.46 |
| ferrite magnet (100 wt % SrM) | 242 | 20 | 24 | 0.31 |

While we acknowledge that it is not straightforward to fully translate this model to the FeCo NWs fabricated in this study that have aspect ratios $L/D$ between 60 (for $D$ = 100 nm) and 200 ($D$ = 30 nm) and present an outer Co ferrite shell, micromagnetic simulations clearly support the hypothesis that the FeCo NWs present a single-domain magnetic configuration that enables the large remanence and competitive coercivity values. In addition, they hint at the formation of the vortex at the core of wires during the magnetization reversal process.

As mentioned in the Introduction section, the two main disadvantages of high-magnetization metallic nanowires as constituents for permanent magnets are as follows: (1) their production usually gives low yields at high cost and (2) their coercivity is not as competitive as that of hard magnetic materials with high magnetocrystalline anisotropy. To circumvent these issues, we fabricate here composites based on NWs as the minority phase and a majority hard ferrite phase. We selected the 50 nm NWs for the composites by discarding the 100 nm FeCo NWs because of their low $H_c$ and the 30 nm ones because of their lower $M_s$.

Figure 5a shows the magnetization curves of oriented SrFe$_{12}$O$_{19}$ (strontium ferrite, SFO)−NW (50 nm) composites with different NW concentrations between 10 and 40 wt %. Figure 5b summarizes the evolution of $H_c$, $M_s$, and $M_r$ as a function of the NW concentration. As expected, given that for SrFe$_{12}$O$_{19}$, $M_s$ = 72 Am$^2$/kg;[51] the $M_s$ of the composites increases, while $H_c$ decreases with the NW content, reaching $M_s$ = 100 Am$^2$/kg and $H_c$ = 104 kA/m for 40 wt %. The identical measurements on composites fabricated with 100 nm NWs presented in Figure S2 of the SI confirm that the NW content needed for a substantial increase in magnetization leads to detrimental coercivity decay.

With the goal of assessing the real potential of these composites as permanent magnets, a real dense composite bonded magnet was fabricated. Based on the results from Figure 5, the composition 70% SrFe$_{12}$O$_{19}$−30 wt % FeCo NW (50 nm) was selected. A SEM image of the composite powder obtained after mixing is shown in Figure 6a, where the nanowires are seen to be dispersed in close proximity with SrFe$_{12}$O$_{19}$ particles. The resulting oriented bonded magnet, schematized on Figure 5b, is thus composed of aligned vertical FeCo NWs embedded in a matrix of oriented SrFe$_{12}$O$_{19}$ (SrM) particles. For comparison, a pure SrM bonded magnet was fabricated in identical conditions to the composite magnet.

In the magnetization curves of Figure 6c, we observe that the composite magnet presents a larger $M_s$ and $M_r$ and a decreased $H_c$. The magnetic parameters are shown in Table 1. Given that the theoretical saturation magnetization for SrM is 72 Am$^2$/kg and the $M_r$ for 50 nm FeCo NW is 120 Am$^2$/kg, the 20% increase in $M_r$ is in agreement with a linear combination of the $M_r$'s of both materials. However, given the existence of magnetostatic interacting fields between all particles, this agreement may just be a coincidence as the exact magnetic state of each phase is unknown. Interwire distances may be larger in the composite with respect to the pure NW sample, which reduces interwire demagnetizing fields, leading to a larger $M_r$ from the NW contribution.[36,37,48] At the same time, these fields may be inducing reversible switching at the zero external field in some of the neighboring ferrite particles, reducing their impact on $M_r$.[33]

This complex nature of the reversal process of magnetic materials, particularly in multiphase systems, also makes $H_c$ values difficult to predict. It is important to remark that the mixing of the two phases was performed by mild sonication, and since no temperature or high-energy milling was employed, we do not expect any exchange coupling to take place between both phases. Nevertheless, based on the coercivity values of the powder composites in Figure 5, the $H_c$ observed in the composite magnet, $H_c$ = 227 kA/m, is much higher than the one measured for the equivalent composite powders, $H_c$ = 131 kA/m. In addition, the bonded composite magnet presents a significant squareness in the magnetization curve, which is not common in hard−soft composites, particularly in the absence of interparticle exchange coupling.

To further understand this behavior, we plot in Figure 6d the derivative of magnetization vs the applied field for the FeCo NW powder, the composite powder, and the bonded magnet. The maxima in the derivative reveal information about the main reversal events in the samples.[52] The pure FeCo NW sample presents a relatively narrow maximum at 36 kA/m, as expected since the narrow distribution of NW diameters and sizes should entail a relatively narrow distribution of coercivities and interaction fields.[33] The oriented composite powder presents two broader maxima, at 57 and 350 kA/m. Given that these values are relatively close to the coercivities of the FeCo NW and the ferrite, respectively, we interpret them as associated with the magnetization reversal of each separate phase. In the bonded magnet, a single broad maximum is observed centered at 215 kA/m, with a small shoulder at 92 kA/m.

The effective anisotropy in a pure NW system is expected to be dominated by the competition between magnetostatic terms: (1) shape anisotropy that favors $M_r$ and $H_c$ and (2) interwire demagnetizing fields that hinder them.[48] An increased packing fraction brings the NW closer and is expected to enhance interwire demagnetizing fields. In the composite magnet, the magnetostatic interactions balance and thus understanding the reversal process is even more complex. On the one hand, the proximity of the platelet-shaped hexaferrite particles modifies the internal fields depending on their exact shape and geometrical arrangement. On the other hand, as explained above, the presence of ferrite particles may be reducing interwire demagnetizing fields with respect to the pure NW sample. More sophisticated characterization methods are needed to fully unravel this balance and the specific reversal mechanisms and switching events.[33,36−39] In any case, the derivatives in Figure 6d suggest that the internal fields and the particle arrangement created by the hard ferrite particles support the initial single-domain state inside the wires, which is likely responsible for the high squareness and $H_c$ of the composite magnet.

Both magnets were measured to have the same density (1.7 g/cm$^3$). Using this value, the energy product (BH$_{max}$) was obtained and is presented in Table 1. A 20% remanence increase in combination with the good squareness of the magnetization curve leads to a BH$_{max}$ value of the FeCo NW composite that is 48% higher than that of the pure ferrite magnet. It is important to remark that the BH$_{max}$ values are considerably lower than those of commercial bonded magnets due to the relatively low densities of the bonded magnets fabricated here. As there was only enough FeCo NW powder for one sample, very mild processing conditions were employed to ensure the survival of the specimen. This led to low densities. The proof of concept is in any case perfectly valid since the reference sample was prepared identically.

From what we learned from comparing the hysteresis loops of composite powders and magnets with equal compositions (Figure 6d), an even greater betterment of the magnetic properties, in this case, $H_c$ and squareness of the loop, would be expected for a composite magnet with a higher density when compared to its corresponding pure SFO reference sample. It is remarkable as well that this increase in magnetic performance is achieved without the need for exchange coupling between both phases, and just as a consequence of the competitive properties of the FeCo NWs and the internal magnetostatic interactions, which simplifies composite fabrication and processing. Future work will include magnetic measurements to further understand the reversal mechanisms and interaction fields inside the composite magnet to optimize particle geometries and arrangements.[38]

## ■ CONCLUSIONS

In summary, we have fabricated high-remanence composites composed of FeCo nanowires and Sr-hexaferrite commercial powders. Soft FeCo NWs with controlled diameters 30, 50, and 100 nm and at least 6 μm long exhibit high remanence under magnetic orientation. Selecting 50 nm as the optimal diameter, a bonded magnet composed of 70 wt % strontium ferrite−30 wt % FeCo NWs is produced and compared to a pure ferrite reference magnet. The observed 20% increase in remanence together with a mild decrease in coercivity, a consequence of the magnetodipolar interaction between ferrite particles and FeCo NWs, yield an enhancement of the energy product of 48% in the composite magnet compared to the pure ferrite magnet. This extremely promising result brings forward nanowire−ferrite composites as a clear candidate for gap magnets that could fill the void between ferrites and rare-earth magnets. The main bottleneck toward industrial implementation at this stage is the low yield and large production cost. Strategies toward solving the issue are being currently investigated.

## ■ METHODS

FeCo NWs were prepared by template-assisted electrochemical deposition using two different nanoporous templates. On the one hand, commercial polycarbonate nanoporous membranes were used, with three different diameters (30, 50, and 100 nm), supplied by Sterlitech. On the other hand, we have fabricated alumina templates, with 50 nm diameter pores, following a one-step anodization process.[31] In this process, high-purity Al foils (99.999%) were anodized at 40 V in 0.3 M $C_2H_2O_4$ solutions at 25° for 7 h. The nanopore diameter was enlarged by etching treatment under phosphoric acid solution (5%) at 50 °C for 40 min.

Electrodeposition was carried out in a three-electrode electrochemical cell in an Ecochemie Autolab PGSTAT potentiostat using a Pt mesh as a counter electrode and an Ag/AgCl (3 M NaCl) electrode as a reference electrode. Before electrodeposition, a thin Au film was thermally evaporated on one side of the membrane to act as a working electrode. The electrolyte was composed of $CoSO_4$ (0.09 M) as a $Co^{2+}$ source, $FeSO_4$ (0.1 M) as an $Fe^{2+}$ source, and $H_3BO_3$ (0.4 M) as an additive. All chemicals were of analytical grade and were used without further purification and mixed in deionized water. The pH was adjusted to 2.7 using 10% vol. $H_2SO_4$. FeCo NWs were grown at RT under a constant voltage of −1.1 V. After growth, the Au layer was removed using a 0.1 M $I_2$ and 0.6 M KI solution. Then, the polycarbonate membrane was dissolved using cycles of sonication in dichloromethane, acetone, and ethanol. Alumina templates were dissolved using a 0.4 M $H_3PO_4$ and 0.2 M $H_2CrO_4$ solution. The released NWs were left to dry at RT in ethanol to recover a dry powder. In addition, to upscale the production yield at lower costs, polycarbonate membranes were substituted by alumina templates with a larger pore density than polycarbonate. Alumina nanoporous templates with a pore of 50 nm were fabricated following a redesigned and faster RT procedure. The details of this process will be published elsewhere. Sr-hexaferrite microsized powders were supplied by Max Baermann GmHb.

The morphology and composition of the NWs were measured with a transmission electron microscope (TEM) JEOL JEM 2000FX. The diameter of the wire and its core were obtained by measuring at three different regions of the wire in the TEM image and by averaging the value. Further structural characterization was performed using secondary electron images from field-emission scanning electron microscopy, field-emission scanning electron microscopy (FE-SEM, Hitachi S-4700), and scanning electron microscopy (TM-1000) at an acceleration voltage set at 15 kV. Structural analysis was performed by X-ray diffraction (XRD) with a Bruker D8 diffractometer using Cu Kα radiation (λ = 1.5418 Å) and a Lynxeye XE-T detector. Subsequent Rietveld analysis of the XRD data was performed by FullProf Suite.[53]

Thermogravimetric analysis (TGA) in air was employed to determine the oxidation stage of the wires between RT and 900 °C using a TA Instruments Q50 system.

For the theoretical study of magnetization reversal in fabricated FeCo NWs, we employ a micromagnetic algorithm initially developed for the simulation of the magnetization distribution of magnetic nanocomposites.[54,55] The magnetization distribution in an isolated cylinder with length $L$ and diameter $D$ under the influence of the external magnetic field is simulated by taking into account the four standard contributions to the total magnetic energy: external field, magnetic anisotropy, exchange, and dipolar interaction. The following material parameters (typical for FeCo alloys) were used for simulating the magnetic nanowires: saturation magnetization, $M_s$ = 220 Am$^2$/kg; cubic magnetocrystalline anisotropy, $K_{cub}$ = 20 kJ/m$^3$ with easy axis along the ⟨100⟩ directions; and exchange stiffness constant, $A_{bulk}$ = 21 pJ/m.[46] NWs under study have the $D$ = 30 nm and $L$ varying in the range 30−400 nm and are discretized in the nonregular mesh with the typical mesh element size of 3 nm. One of the cubic crystallographic axes and the direction of the external magnetic field coincide with the FeCo NW rotation axis; the directions of the other two axes are identical for all mesh elements. Open boundary conditions were applied in all simulations. Note that every set of parameters employs different irregular mesh realizations, resulting in slightly different coercive fields.

FeCo NW−strontium ferrite mixture composites in the powder form were fabricated by mixing a few milligrams of each phase at 10, 20, 30, and 40 wt % of FeCo NWs and sonicating in an ethanol bath for 5 min. The solution was left to dry and the composite powder was recovered (approximate mass around 1−3 mg).

To fabricate a consolidated bonded magnet, 90 mg of NW ($D$ = 50 nm) powder was fabricated and mixed by sonication in ethanol with 210 mg of strontium ferrite ($SrFe_{12}O_{19}$). The resulting composite powder was dried and added to a mix of epoxy resin and hardener in a 1:1 ratio. The mixture was homogenized and added to a mold while applying an external magnetic field of 0.4 T to orient the magnet. Subsequently, constant uniaxial pressure was applied for 24 h. On the other hand, using 300 mg of ferrite powder, a pure ferrite reference bonded magnet was fabricated for comparison purposes.

Magnetic characterization of the powders was carried out at RT by means of a Lakeshore vibrating sample magnetometer (VSM), with a maximum applied field of 1.8 T, and a Quantum Design MPMS SQUID magnetometer with a maximum applied field of 5 T. Each of the nanowire powders, corresponding to the three different diameters, and the resulting composites with strontium ferrite were individually dispersed inside a bonding glass matrix (509 Crystalbond) and left to consolidate under an applied magnetic field. Per our alignment procedure, the mass of powder introduced in the bonding glass cannot be accurately determined. For this reason, nonoriented samples were measured to calibrate the magnetization per sample mass. Besides, a homemade VSM, whose maximum magnetic field is 1.3 T, was used to characterize the magnetic properties of the bonded magnets.[56]

## ASSOCIATED CONTENT

### Supporting Information

The Supporting Information is available free of charge at https://pubs.acs.org/doi/10.1021/acsanm.0c01905.

SEM images of the 30 and 100 nm diameter NW powders (Figure S1), magnetization curves of ferrite−NW composites made with 100 nm diameter FeCo NWs (Figure S2), and magnetization curves of the 50 nm FeCo NWs measured several days and 1 year after exposure to air (PDF)

## AUTHOR INFORMATION

### Corresponding Author

A. Quesada − *Instituto de Cerámica y Vidrio (CSIC), Madrid 28049, Spain;* orcid.org/0000-0002-6994-0514; Email: a.quesada@icv.csic.es

### Authors

J. C. Guzmán-Mínguez − *Instituto de Cerámica y Vidrio (CSIC), Madrid 28049, Spain;* orcid.org/0000-0003-1017-0225

S. Ruiz-Gómez − *Departamento de Física de Materiales, Universidad Complutense de Madrid, Madrid 28040, Spain*

L. M. Vicente-Arche − *Instituto de Cerámica y Vidrio (CSIC), Madrid 28049, Spain; Unité Mixte de Physique, CNRS, Thales, Université Paris-Saclay, 91767 Palaiseau, France*

C. Granados-Miralles − *Instituto de Cerámica y Vidrio (CSIC), Madrid 28049, Spain;* orcid.org/0000-0002-3679-387X

C. Fernández-González − *IMDEA Nanociencia, 28049 Madrid, Spain*

F. Mompeán − *Instituto de Ciencia de Materiales de Madrid (CSIC), Madrid 28049, Spain*

M. García-Hernández − *Instituto de Ciencia de Materiales de Madrid (CSIC), Madrid 28049, Spain*

S. Erohkin − *General Numerics Research Lab, 07745 Jena, Germany*

D. Berkov − *General Numerics Research Lab, 07745 Jena, Germany*

D. Mishra − *Institute of Materials for Electronics and Magnetism-CNR, 43124 Parma, Italy; Department of Physics, Indian Institute of Technology Jodhpur, Jodhpur 342037, Rajasthan, India*

C. de Julián Fernández − *Institute of Materials for Electronics and Magnetism-CNR, 43124 Parma, Italy;* orcid.org/0000-0002-6671-2743

J. F. Fernández − *Instituto de Cerámica y Vidrio (CSIC), Madrid 28049, Spain;* orcid.org/0000-0001-5894-9866

L. Pérez − *Unité Mixte de Physique, CNRS, Thales, Université Paris-Saclay, 91767 Palaiseau, France; IMDEA Nanociencia, 28049 Madrid, Spain;* orcid.org/0000-0001-9470-7987

Complete contact information is available at:
https://pubs.acs.org/10.1021/acsanm.0c01905

### Notes
The authors declare no competing financial interest.

## ACKNOWLEDGMENTS

We would like to thank Dr. Víctor Fuertes for his advice on the processing of the bonded magnets. This work is supported by the Spanish Ministerio de Economía y Competitividad y Ministerio de Ciencia e Innovación (Project Nos. MAT2017-86450-C4-1-R, MAT2015-64110-C2-1-P, MAT2015-64110-C2-2-P, MAT2017-87072-C4-2-P, RTI2018-095303-A-C52, and FIS2017-82415-R) and by the European Commission through Project H2020 (No. 720853; AMPHIBIAN). C.G.-M. acknowledges financial support from MICINN through the "Juan de la Cierva" Program (FJC2018-035532-I). A.Q. acknowledges financial support from MICINN through the "Ramón y Cajal" Program (RYC-2017-23320). The work also is funded by the Regional Government of Madrid (Project S2018/NMT-4321; NANOMAGCOST).